\documentclass[11pt]{article}
\usepackage{amsmath,amssymb,color,epsfig,cite}
\usepackage{graphicx}
\usepackage{setspace}

\usepackage{amsfonts}
\newcommand{\hoch}[1]{$\, ^{#1}$}

\makeatletter
\@addtoreset{equation}{section}
\makeatother

\newcommand{\be}{\begin{equation}}
\newcommand{\ee}{\end{equation}}
\newcommand{\bea}{\setlength\arraycolsep{2pt} \begin{eqnarray}}
\newcommand{\eea}{\end{eqnarray}}

\newcommand{\half}{{\textstyle{\frac{1}{2}}}}

\def\ndelta{\delta\hspace{-0.50em}\slash\hspace{-0.05em} }

\def\ft#1#2{{\textstyle{\frac{\scriptstyle #1}{\scriptstyle #2} } }}
\def\fft#1#2{{\frac{#1}{#2}}}

\def\0{{\sst{(0)}}}
\def\1{{\sst{(1)}}}
\def\2{{\sst{(2)}}}
\def\3{{\sst{(3)}}}
\def\4{{\sst{(4)}}}
\def\5{{\sst{(5)}}}
\def\6{{\sst{(6)}}}
\def\7{{\sst{(7)}}}
\def\8{{\sst{(8)}}}
\def\sst#1{{\scriptscriptstyle #1}}

\def\del{{\partial}}

\begin{document}

\begin{flushright}
\hfill{\hfill{MI-TH-1890}}

\end{flushright}

\vspace{15pt}
\begin{center}
{\Large {\bf Subleading BMS charges and fake news near null infinity}}

\vspace{15pt}
{\bf Hadi Godazgar\hoch{1}, Mahdi Godazgar\hoch{2} and 
C.N. Pope\hoch{3,4}}

\vspace{10pt}

\hoch{1} {\it Max-Planck-Institut f\"ur Gravitationsphysik (Albert-Einstein-Institut), \\
M\"uhlenberg 1, D-14476 Potsdam, Germany.}

\vspace{10pt}

\hoch{2} {\it Institut f\"ur Theoretische Physik,\\
Eidgen\"ossische Technische Hochschule Z\"urich, \\
Wolfgang-Pauli-Strasse 27, 8093 Z\"urich, Switzerland.}

\vspace{10pt}

\hoch{3} {\it George P. \& Cynthia Woods Mitchell  Institute
for Fundamental Physics and Astronomy,\\
Texas A\&M University, College Station, TX 77843, USA.}

\vspace{10pt}

\hoch{4}{\it DAMTP, Centre for Mathematical Sciences,\\
 Cambridge University, Wilberforce Road, Cambridge CB3 OWA, UK.}


\vspace{20pt}

\underline{ABSTRACT}
\end{center}

\noindent 
In this paper we establish a 
relation between the non-linearly conserved Newman-Penrose charges
and certain subleading terms in a large-$r$ expansion of the BMS charges
in an asymptotically-flat spacetime.
We define the subleading BMS charges by considering a $1/r$-expansion of the 
Barnich-Brandt prescription for defining asymptotic charges in an 
asymptotically-flat spacetime.  At the leading order, i.e.~$1/r^0$, one
obtains the standard BMS charges, which would be integrable and conserved
in the absence of a flux term at null infinity, corresponding to gravitational 
radiation, or Bondi news.  At subleading orders, analogous terms 
in general provide obstructions to the integrability of the corresponding 
charges.  Since 
the subleading terms are defined close to null infinity, but vanish actually
at infinity, the analogous obstructions are not associated with genuine 
Bondi news.  One may instead describe them as corresponding to ``fake news.'' 
At order $r^{-3}$, we find that a set of integrable charges can be 
defined and that these are related to the ten non-linearly
conserved Newman-Penrose charges.

\thispagestyle{empty}

\vfill
E-mails: hadi.godazgar@aei.mpg.de, godazgar@phys.ethz.ch, pope@physics.tamu.edu

\pagebreak

\section{Introduction}

The asymptotic symmetry group of asymptotically-flat spacetimes, 
the BMS group and its associated charges, has encountered somewhat of a 
resurgence in interest recently, whether in the context of flat space holography \cite{BarnichAspects, Barnich:2013axa}, its relation to the Weinberg soft theorems \cite{Strominger:2013jfa, He:2014laa} or to black holes physics \cite{Hawking:2016msc,Hawking:2016sgy,Sheikh-Jabbari:2016lzm}.  

The novel feature of asymptotically-flat spacetimes is that their asymptotic 
symmetry group \cite{bondi, sachs} as one asymptotically approaches null 
infinity is much larger than the na\"ively expected Poincar\'e group, 
the symmetry group of Minkowski spacetime.  It is the existence of an 
infinite number of supertranslations that distinguishes the BMS group from the 
Poincar\'e group.  More precisely, the BMS group is the semi-direct 
product of conformal isometries on the round 2-sphere with the
supertranslations, i.e.~angle-dependent translations along future null
infinity (see equation (\ref{BMSgen})):
\begin{equation}
\textup{BMS} = \textup{SL$(2,\mathbb{C})$} \ltimes \textup{ST}.
\end{equation}
Whether viewed from a 
phase-space \cite{Ashtekar:1981bq, LW, IW, WZ} or covariant \cite{BB,BarTro}  
point of view, the existence of an enhanced (infinite) asymptotic symmetry 
group implies the existence of an infinite number of charges; the BMS charges. 
 Roughly speaking, the BMS charges are constructed by integrating a 
BMS transformation parameter multiplied by a BMS invariant quantity over 
the sphere at null infinity.  Of course, in the non-linear theory there is 
the subtle issue that charges will generally not be integrable due to 
the existence of flux at infinity, associated with gravitational
radiation (measured by the Bondi flux, or Bondi news) \cite{bondi, WZ, BarTro}.

A short time after the BMS group and its associated charges were discovered, 
another set of (conserved) charges at null infinity was also discovered, 
known as Newman-Penrose (NP) charges \cite{NP}.  Newman and Penrose 
constructed their charges in the framework of the Newman-Penrose 
formalism \cite{NP61}.  These charges \emph{are} conserved along null 
infinity, and are given by the integral over the sphere at infinity of a 
particular spherical harmonic of a Weyl scalar. In the linearised theory there 
is an infinite tower of such charges, while in the non-linear theory the 
tower collapses to ten such NP charges.  Despite the fact that the existence of NP charges requires a leading analytic expansion for the fields around null infinity, which is in general not satisfied \cite{Damour:1985cm, christ}, NP charges have also been of interest recently in relation to the existence of conserved charges on the horizon of extremal back holes \cite{Aretakis:RN1, Aretakis:extremal, BF, LuciettiERN, us}.  In Ref.\ \cite{us}, it has been shown that there is a 1-1 correspondence between Aretakis charges on the extremal horizon and NP charges at null infinity of so-called weakly asymptotically-flat spacetimes.

The question that we would like to address here is the relation between 
BMS and NP charges. 

At first glance there is no obvious relation between these two sets of 
charges, but, given that they are both defined in the asymptotic region of 
asymptotically-flat spacetimes, it would seem natural that there
should exist some connection between them.  For simplicity, we shall 
restrict our attention henceforth to the supertranslations.  
Generalising to the full BMS group 
should not be too difficult.  However, since the most interesting part is 
the supertranslations, it makes sense to focus our attention on these 
transformations.

Recently, it was shown by Conde and Mao in Ref.\ \cite{conde} that in the 
linearised theory the infinite tower of NP charges may be reinterpreted as 
subleading BMS charges.  The standard BMS charge associated with 
supertranslations is given by the integral over the sphere at infinity of 
the Bondi mass aspect, which is supertranslation invariant in the 
linearised 
theory, multiplied by a supertranslation parameter.  What Conde and Mao 
realised is that the Bondi mass aspect is but the leading $1/r^{0}$ term in a 
$1/r$-expansion of the $uu$-component of the linearised metric 
perturbation $\delta g_{ab}$.  Furthermore, $\delta g_{uu}$ is invariant 
under supertranslations.  This led them to define a new BMS charge at each 
order in the $1/r$-expansion, finding that the subleading BMS charges 
include the infinite tower of NP charges that exist in the linear 
theory.~\footnote{In fact they only identify the real part of the NP 
charges, because their expansion for the BMS charge is real. We shall 
encounter the same feature in the non-linear case.}

Our aim in this paper is to generalise the above result to the full 
non-linear theory.  As pointed out before, this is non-trivial given the 
existence of flux in the non-linear theory.  
In particular,  $\delta g_{uu}$ is no longer supertranslation invariant. Moreover, generally, in the non-linear theory the objects of interest are not supertranslation invariant. Hence, the same method as Conde-Mao cannot be used to find the non-linear charges.
Our idea is very simple:  
we take as our starting point the general expression for asymptotic charges 
derived by Barnich and Brandt \cite{BB}.~\footnote{There is an ambiguity in the definition of the asymptotic charges in general relativity (see Ref.~\cite{compere} for a discussion of this point).  However, this ambiguity will not affect the results in this paper (see section \ref{sec:dis} for more details).}  As defined, the Barnich-Brandt 
expression can be considered as a $1/r$-expansion, the leading $1/r^{0}$ term being 
the standard BMS charge. Thus, each subsequent term in this $1/r$-expansion 
may be viewed as a subleading BMS charge.  We find that at 
order $r^{-3}$, the subleading BMS charges are associated with
the non-linearly conserved NP charges.

We begin in section \ref{sec:AF} by reviewing properties of asymptotically-flat spacetimes, as defined by Bondi \cite{bondi}.  We explain the fall-off conditions that will be assumed in this paper, the canonical complex null frame for the general metric, the form of the Einstein equations at each order and, most importantly, the BMS group and how it acts on the fields.

In section \ref{sec:BMSsub}, we consider a $1/r$-expansion of the 
Barnich-Brandt definition of the asymptotic charge adapted to 
asymptotically-flat spacetimes, defining these to be subleading BMS charges.  
We analyse the expansion up to order $r^{-3}$.  In general, the structure 
of the subleading BMS charges is similar to that of the leading charges; 
there exist both integrable and non-integrable pieces.  At each order, 
we consider whether the non-integrable pieces can be made to vanish by
making particular choices for the supertranslation parameter, 
finding that this can only be done non-trivially at order $r^{-3}$.

The relation of the subleading BMS charges to the Newman-Penrose formalism 
is clarified in section \ref{sec:BMSNP}.  In particular, we show that the 
integrable BMS charges at order $r^{-3}$ correspond to NP charges.  
We conclude with some comments in section \ref{sec:dis}.

\section{Asymptotically-flat metrics} \label{sec:AF}

Here, we work with the Bondi definition of asymptotic flatness \cite{bondi, sachs}.  We introduce Bondi coordinates $(u,r,x^I=\{\theta,\phi\})$, such that the metric takes the form
\begin{equation} \label{AF}
 d s^2 = - F e^{2 \beta} du^2 - 2 e^{2 \beta} du dr + 
r^2 h_{IJ} \, (dx^I - C^I du) (dx^J - C^J du)
\end{equation}
with the metric functions satisfying the following fall-off conditions at large $r$
\begin{align}
 F(u,r,x^I) &= 1 + \frac{F_0(u,x^I)}{r} + \frac{F_1(u,x^I)}{r^2} + \frac{F_2(u,x^I)}{r^3} + \frac{F_3(u,x^I)}{r^4} + o(r^{-4}), \notag \\[2mm]
 \beta(u,r,x^I) &= \frac{\beta_0(u,x^I)}{r^2} + \frac{\beta_1(u,x^I)}{r^3} + \frac{\beta_2(u,x^I)}{r^4} + o(r^{-4}), \notag \\[2mm] 
 C^I(u,r,x^I) &= \frac{C_0^I(u,x^I)}{r^2} + \frac{C_1^I(u,x^I)}{r^3} + \frac{C_2^I(u,x^I)}{r^4} + \frac{C_3^I(u,x^I)}{r^5} + o(r^{-5}), \notag \\[2mm] \label{met:falloff}
 h_{IJ}(u,r,x^I) &= \omega_{IJ} + \frac{C_{IJ}(u,x^I)}{r} + \frac{C^2 \omega_{IJ}}{4 r^2} + \frac{D_{IJ}(u,x^I)}{r^3} + \frac{E_{IJ}(u,x^I)}{r^4} + o(r^{-4}),
\end{align}
where $\omega_{IJ}$ is the standard metric on the round 2-sphere with coordinates $x^I=\{\theta, \phi\}$ and  $C^2 \equiv C_{IJ} C^{IJ}$.  Moreover, residual gauge freedom allows us to require that
\begin{equation} \label{det:h}
 h =\omega,
\end{equation}
where $h \equiv \textup{det}(h_{IJ})$ and $\omega 
                      \equiv \textup{det}(\omega_{IJ}) =\sin\theta$.
A parameterisation of $h_{IJ}$, which makes this gauge choice obvious is one for which \cite{sachs}
\begin{equation}
 2 h_{IJ} dx^I dx^J = (e^{2f} + e^{2g}) d\theta^2 + 4 \sin{\theta} \sinh(f-g) d\theta d\phi + \sin^2\theta (e^{-2f} + e^{-2g}) d\phi^2
\end{equation}
with
\begin{align}
  f(u,r,x^I) &= \frac{f_0(u,x^I )}{r}+\frac{f_2(u,x^I)}{r^3} +\frac{f_3(u,x^I)}{r^4} + o(r^{-4}), \notag \\[1mm]
  g(u,r,x^I) &= \frac{g_0(u,x^I)}{r}+\frac{g_2(u,x^I)}{r^3} +\frac{g_3(u,x^I)}{r^4} + o(r^{-4}). \label{def:fg}
 \end{align}
 Note that there are no terms above for $f$ and $g$ at order $r^{-2}$ because of regularity conditions on the metric \cite{sachs}.
 
As will become clear later, both parameterisations for $h_{IJ}$ are useful 
and, clearly, there is a relation between the two.  In particular, we have
\begin{gather}
 C_{IJ} = \text {{\footnotesize $ \begin{pmatrix}
            f_0 + g_0 & (f_0 - g_0) \sin \theta \\
            (f_0 - g_0) \sin \theta & -(f_0 + g_0) \sin^2 \theta
          \end{pmatrix}$} }, \quad
 D_{IJ} = \text {{\footnotesize $ \begin{pmatrix}
            f_2 + g_2 + \ldots & (f_2 - g_2  + \ldots) \sin \theta  \\
            (f_2 - g_2  + \ldots) \sin \theta & -(f_2 + g_2  + \ldots) \sin^2 \theta
          \end{pmatrix}$} }, \notag \\[2mm]
 E_{IJ} = \text {{\footnotesize $ \begin{pmatrix}
            f_3 + g_3 + \ldots & (f_3 - g_3  + \ldots) \sin \theta  \\
            (f_3 - g_3  + \ldots) \sin \theta & -(f_3 + g_3  + \ldots) \sin^2 \theta
          \end{pmatrix}$} },         
\end{gather}
where the ellipses indicate lower order terms in $f$ and $g$, such as $f_0$ and $g_0$.

Since we are using the gauge \eqref{det:h} in which the determinant of 
$h_{IJ}$ is equal to the  determinant of the round metric on the 
2-sphere, this implies that $C_{IJ}$ and $D_{IJ}$ are both trace-free, while
\begin{equation} \label{trE}
\textup{tr}\, E \equiv \omega^{IJ} E_{IJ} = D^{IJ} C_{IJ} - \frac{1}{16} \left(C^2 \right)^2,
\end{equation}
where
\begin{equation}
 C^2 \equiv C_{IJ} C^{IJ} = 4 (f^2_0 + g^2_0).
\end{equation}

\subsection{Null frame} \label{sec:frame}

A complex null frame $e_\mu{}^a=(\ell^a,n^a,m^a,\bar{m}^a)$ with inverse $E^\mu{}_a$,
\begin{equation}
 g_{ab} = E^\mu{}_a E^\nu{}_b \ \eta_{\mu \nu}, \qquad \eta_{\mu \nu} = \text {{\footnotesize $ \begin{pmatrix}
                                                                       \begin{matrix} 0 & -1 \\ -1 & 0 \end{matrix} & \mathbf{0} \\
                                                                       \mathbf{0} & \begin{matrix} 0 & 1 \\ 1 & 0 \end{matrix}
                                                                    \end{pmatrix}$ }}
\end{equation}
may be introduced, where
\begin{align}
 \ell &= \frac{\partial}{\partial r}, \qquad n =  e^{- 2 \beta} \Bigg[ \frac{\partial}{\partial u} - \half F \frac{\partial}{\partial r} + C^I \frac{\partial}{\partial x^I} \Bigg], \qquad m = \frac{\hat{m}^I}{r}  \frac{\partial}{\partial x^I}, \notag\\
 \ell^\flat &= - e^{2\beta} du, \qquad n^{\flat} = - \Big( dr + \frac{1}{2} F du \Big), \qquad m^{\flat} = r\, \hat{m}_I\, (dx^I - C^I du),
 \label{AF:frame}
\end{align}
where 
\begin{equation}
 2 \hat{m}^{(I} \bar{\hat{m}}^{J)} = h^{IJ}
\end{equation}
with $h^{IJ}$ the matrix inverse of $h_{IJ}$.  Equivalently,
\begin{equation}
 m = \frac{1}{2r} \left[ (e^{-f} + i e^{-g}) \partial_\theta -\frac{i}{\sin\theta} (e^{f} + i e^{g}) \partial_\phi \right].
\end{equation}
Given some arbitrary vector $V_a$, we denote the components in the null basis as follows
\begin{equation}
 \ell^a V_a \equiv V_0 = - V^1,\qquad n^a V_a \equiv V_1 = -V^0,\qquad m^a V_a \equiv V_m=V^{\bar{m}},
\end{equation}
with the obvious generalisation also to tensors.

\subsection{Einstein equations} 

As well as the fall-off conditions \eqref{met:falloff} and the gauge condition \eqref{det:h}, following Ref.\ \cite{sachs}, we assume that the components 
$T_{00}$ and $T_{0m}$ of the energy-momentum tensor in the null frame 
fall off as
\begin{equation} \label{falloff:matter}
 T_{00} = o(r^{-5}), \qquad T_{0m} = o(r^{-3}).
\end{equation}
The Einstein equation then implies that
\begin{align}
  G_{00} = o(r^{-5}) &\quad \implies \quad \beta_0 = -\frac{1}{32}\, C^2, \quad \beta_1 = 0,  \\
  G_{0m} = o(r^{-3}) &\quad \implies \quad C_0^I = -\half D_J C^{IJ},
\end{align}
where $D_I$ is the standard covariant derivative associated 
with the round-sphere metric $\omega_{IJ}$.

Furthermore, at higher orders, given appropriate fall-off for energy-momentum tensor components, the Einstein equation would imply the following equations
\begin{align}
  G_{00} = o(r^{-6}) &\ \implies \ \beta_2 = -\frac{3}{32} D_{IJ} C^{IJ} + \frac{1}{128}\, (C^2)^2, \label{b2} \\[2mm]
  G_{0m} = o(r^{-5}) &\ \implies \ C_2^I = \frac{3}{4} \left( D_J D^{IJ} - C^{IJ} C_{1\, J} \right) +\frac{1}{64} C^2 D_J C^{IJ} -\frac{1}{16} C^{IJ} D_{J}C^2, \label{C2} \\[2mm]
  G_{0m} = o(r^{-6}) &\ \implies \ C_3^I = \frac{2}{5} D_{J} E^{IJ} + \frac{9}{80} C^2 C_1^I -\frac{19}{80} C_{KL} D^K D^{LI} -\frac{51}{80} C^{IL} D^K D_{KL} \notag \\
                                            & \hspace{22mm} -\frac{11}{80} D^{KL} D^I C_{KL} + \frac{7}{160} C^2 D^I C^2, \label{C3}
\end{align}
\begin{align}
  G_{mm} = o(r^{-4}) &\ \implies \ \partial_u D_{IJ} = \frac{1}{8} C_{IJ} \partial_u C^2 - \frac{1}{4} F_{0} C_{IJ}  - \frac{1}{2} D_{(I} C_{1\, J)} - \frac{1}{8} C_{IJ} D_K D_L C^{KL} \notag \\
            & \hspace{8mm} + \frac{1}{32} D_I D_J C^2 + \frac{1}{2} D_{(I}(C_{J)K} D_L C^{KL}) - \frac{1}{8} D_I C^{KL} D_J C_{KL}    \notag \\
            & \hspace{8mm} + \frac{1}{4} \omega_{IJ} \Big[ D_K C_1^K -\frac{5}{16} \Box C^2  + D^M C^{KL} \big( D_{K} C_{LM}- \frac{1}{4} D_{M} C_{KL}\big) + C^2 \Big], \label{uD} \\[2mm]
  G_{mm} = o(r^{-5}) &\ \implies \ \partial_u E_{IJ} = \frac{1}{2} D^K(C_{1\, (I} C_{J)K}) - \frac{1}{2} D^K D_{(I} D_{J)K} + \frac{5}{32} D^K(C^2 D_{(I} C_{J)K})  \notag \\
            & \hspace{8mm}  - \frac{1}{8} D^K (C_{K(I} D_{J)} C^2) + \frac{1}{2} \omega_{IJ} \Big[ D^{KL} \partial_u C_{KL} - \frac{1}{4} C^2 F_0 - \frac{1}{2} C_{1}^{K} D^L C_{KL}    \notag \\            
            & \hspace{8mm}  - C^{KL} D_K C_{1\, L}+ \frac{1}{2} D^K D^L D_{KL} - \frac{1}{32} C^2 D^K D^L C_{KL} + \frac{5}{32} C^{KL} D_K D_L C^2  \notag \\
            & \hspace{8mm}  - \frac{1}{16} C_{KL} D_M C^{MK} D_N C^{NL} + \frac{3}{32} C^{KL} D_K C^{MN} D_L C_{MN} \Big],  \label{uE}
\end{align}
\begin{align}
  G_{01} = o(r^{-4}) &\ \implies \ F_1 = -\frac{1}{2} D_I C_1^I + \frac{3}{32} (\Box - 2) C^2 \notag \\
                                                                      & \hspace{45mm}         + \frac{1}{2} D_{I} C^{IK} D^J C_{JK} -\frac{1}{8} D^{I} C^{JK} D_{I}C_{JK}, \label{F1} \\[2mm]
  G_{01} = o(r^{-5}) &\ \implies \ F_2 = -\frac{1}{4} D_I D_J D^{IJ} -\frac{3}{4} C_1^I D^J C_{IJ} + \frac{1}{32} C^{IJ} C^{KL} \, D_I D_J C_{KL}  \notag \\
                                                & \hspace{20mm} + \frac{1}{64} C^2 \, D_I D_J C^{IJ} - \frac{1}{32} C^{IJ} D_I C^{KL} D_J C_{KL} + \frac{5}{64} D_I C^{IJ} D_J C^{2}, \label{F2} 
\end{align}
\begin{align}                                               
  G_{01} = o(r^{-6}) &\ \implies \ F_3 = -\frac{1}{10} D_I D_J E^{IJ} + \frac{3}{4} C_1^I C_{1\, I} + \frac{3}{160} D_I(C^2 C_1^I) +\frac{5}{512} (C^2)^2 \notag \\
            & \hspace{20mm} +\frac{1}{16} C^{IJ} \Box D_{IJ} + \frac{9}{80} D^{IJ} \Box C_{IJ} -\frac{11}{40} D^I C^{JK} D_I D_{JK}   \notag \\
            & \hspace{20mm} + \frac{2}{5} D^I C^{JK} D_J D_{IK} - \frac{3}{80} D^{IJ} C_{IJ}-\frac{33}{5120} \Box (C^2)^2 \notag \\
            & \hspace{20mm} + \frac{13}{1024} D^I C^2 D_I C^2 + \frac{3}{128} C^2 D^{I} C^{JK} D_I C_{JK} \notag \\
            & \hspace{20mm} - \frac{1}{32} C^2 D^{I} C^{JK} D_{J} C_{IK}, \label{F3}
\end{align}
\begin{align}            
  G_{11} = o(r^{-2}) &\ \implies \ \partial_u F_0 = -\frac{1}{2} D_I D_J \partial_u C^{IJ} + \frac{1}{4} \partial_u C^{IJ} \partial_u C_{IJ}, \label{uF0} \\[2mm]
  G_{1m} = o(r^{-3}) &\ \implies \ \partial_u C_1^I = \frac{1}{3} D^I F_0 +\frac{1}{6} \Box D_J C^{IJ} - \frac{1}{6} D^I D^J D^K C_{JK} + \frac{1}{8} C_{JK} \partial_u D^I C^{JK} \notag \\
  & \hspace{25mm} + \frac{5}{8} \partial_u C_{JK} D^{I} C^{JK} -\frac{2}{3} \partial_u C_{JK} D^{J} C^{KI} -\frac{1}{6} D_J C^{IJ}, \label{uC1}
\end{align}
where $\Box \equiv D^I D_I$ is the covariant Laplacian on the unit 2-sphere.

\subsection{BMS group} \label{sec:BMS}

The asymptotic BMS symmetry is determined by imposing that the variation of the metric under the generators of the asymptotic symmetry group respects the form of the metric and the gauge choices.
These conditions imply that~\footnote{As explained in the introduction, for simplicity, we neglect the SL$(2,\mathbb{C})$ part of the BMS group.}
\begin{equation} \label{BMSgen}
 \xi = s\, \partial_u +   \int dr \frac{e^{2\beta}}{r^2} h^{IJ} D_{J} s \  \partial_I - \frac{r}{2} \left( D_I \xi^I - C^I D
 _I s \right) \partial_r.
\end{equation}
The $u$ and $r$-independent function $s(x^I)$ parameterises supertranslations.

We list below the variation of some of the metric components under supertranslations that will be useful later.  Some of these variations can also be found in Ref. \cite{BarTro}.  
\begin{align}
 \delta F_0 &= s \partial_u F_0 - \frac{1}{2} \partial_u C^{IJ} D_I D_J s -  D_I \partial_uC^{IJ} D_J s, \label{var:F0} \\[2mm]
 \delta C_1^I &= s \partial_u C_1^I + \frac{1}{16} \partial_u C^2 D^I s + F_0 D^I s - \frac{1}{4} C^{JK} D^I D_J D_K s - \frac{1}{2} C^{IJ} D_J \Box s 
 \notag \\
 &+ \frac{1}{2} D^J C^{IK} D_J D_K s - \frac{3}{4} D^I C^{JK} D_J D_K s - \frac{1}{2} D_J C^{JK} D_K D^I s - \frac{1}{2} D^I D^J C_{JK} D^K s \notag \\
 &+ \frac{1}{2} D^J D_K C^{KI} D_J s - C^{IJ} D_J s, \label{var:C1} 
 \end{align}
 \begin{align}
 \delta C_{IJ} &= s \partial_u C_{IJ} + \Box s\ \omega_{IJ} - 2 D_{(I} D_{J)} s, \label{var:C} \\[2mm]
 \delta C^2 &= s \partial_u C^2 - 4 C^{IJ} D_I D_J s, \label{var:C2} \\[2mm]
 \delta D_{IJ} &= s \partial_u D_{IJ} + \Big[ \frac{1}{16} C^2 \Box s - \frac{1}{16} D^K C^2 D_K s - \frac{1}{2} C^{LM} D^K C_{KL} D_Ms + C_1^K D_K s \Big] \omega_{IJ}  \notag \\
 & - 2 C_{1 \, (I} D_{J)}s- \frac{1}{4} C_{IJ} C^{KL} D_K D_L s  - \frac{1}{8}C^2 D_{I} D_{J}s + \frac{1}{8} D_{(I} C^2 D_{J)}s + D_K C^{KL} C_{L(I} D_{J)}s, \label{var:D}  \\[2mm]
 \delta E_{IJ} & = s \partial_u E_{IJ}  + \Big[ \frac{1}{4} D^{KL} D_K D_L s + \frac{3}{2} D_K D^{KL} D_L s  - \frac{5}{4}  C^{KL} C_{1\, K} D_L s - \frac{1}{64}  C^2 C^{KL} D_K D_L s  \notag \\ 
 & + \frac{3}{64} \Big( C^{KL} D_K C^2+ 2 C^2 D_K C^{KL} \Big) D_L s\Big] \omega_{IJ} + \frac{1}{2} C_{1\, (I} C_{J)K} D^K s - \frac{5}{2} D^K( D_{K(I} D_{J)} s) \notag \\ 
 & - \frac{1}{2} D^K s D_{(I} D_{J)K}  + \frac{5}{32} D^K (C^2 C_{K(I} D_{J)} s) + \frac{5}{32} C^2 D^K s D_{(I} C_{J)K} - \frac{1}{8} C_{K(I} D_{J)}C^2 D^K s. \label{var:E}
\end{align}

As explained above, the form of the Bondi metric \eqref{AF} is preserved under the action of the BMS group.  However, assuming a particular fall-off for the energy-momentum tensor components implies, via the Einstein equations, additional constraints on the metric.  Of course, one must be sure that these extra conditions are also preserved under the action of the symmetry group.  They will be preserved as long as a particular set of energy-momentum tensor components satisfy particular fall-off conditions.  More precisely, consider the variation of a particular component
\begin{equation} \label{var:Einstein}
 \delta_{\xi} T_{\alpha \beta} = (\mathcal L_{\xi} T)_{\alpha \beta} = \xi^c \partial_c T_{\alpha \beta} + T_{c \beta} \partial_\alpha \xi^c + T_{\alpha c} \partial_\beta \xi^c,
\end{equation}
where $\alpha$ and $\beta$ denote a fixed component of $T_{ab}$ in the null frame, i.e.\ they are each chosen from the set $\{0,1,m,\bar{m}\}$.  
Now, assuming that
\begin{equation}
 T_{\alpha \beta} = o(r^{-n}),
\end{equation}
for some integer $n$, equation \eqref{var:Einstein} at $O(r^{-n})$ equals
\begin{equation}
 \delta_{\xi} T_{\alpha \beta} = T_{c \beta} \partial_\alpha \xi^c + T_{\alpha c} \partial_\beta \xi^c.
\end{equation}
Therefore, a necessary condition that the fall-off condition for $T_{\alpha \beta}$ be preserved is that $T_{c \alpha}$ and $T_{c \beta}$ also satisfy appropriate fall-off conditions.  Here, when assuming a particular fall-off condition for a particular component of $T_{ab}$, we will always assume that the relevant components of $T_{ab}$ also satisfy appropriate fall-off conditions such that the fall-off condition for $T_{\alpha \beta}$ is preserved by the action of the BMS group.  This can always be done.

\section{BMS charges at subleading order} \label{sec:BMSsub}

An expression for the variation of an asymptotic charge in general relativity is given by Barnich and Brandt \cite{BB} (see also Ref.~\cite{Abbott:1981ff})
\begin{gather} 
  \ndelta \mathcal{Q}_\xi[\delta g, g]= \frac{1}{8 \pi G} \int_{S}\,(d^2x)_{ab}\, \sqrt{-g}\ \Big\{ \xi^b g^{cd} \nabla^a \delta g_{cd} -\xi^b g^{ac} \nabla^d \delta g_{cd} +\xi^c g^{ad} \nabla^b \delta g_{cd} \hspace{20mm} \notag   \\[2mm]
  \hspace{70mm} + \frac{1}{2} g^{cd} \delta g_{cd} \nabla^b \xi^a + \frac{1}{2} g^{bd} \delta g_{cd} (\nabla^a \xi^c - \nabla^c \xi^a) \Big\} , \label{AsympCharge}
\end{gather}
where
\begin{equation}
 (d^2x)_{ab} = \frac{1}{4} \eta_{abIJ}\ d x^I \wedge d x^J,
\end{equation}
where $\eta$ is the alternating symbol with $\eta_{u r \theta \phi}=1$. The
slash on the variational symbol $\delta$ signifies the fact that the
variation is not, in general, integrable. 

As is explained in section \ref{sec:dis}, the above definition is not unique.  For example, it differs from the expression given by Iyer and Wald by an ambiguity, which vanishes for $\xi$ an exact Killing vector, as opposed to an asymptotic one.  We find that the ambiguity vanishes also in this case, rendering all such charges equal.
 
The background of interest here, with metric $g_{ab}$, is the class of asymptotically-flat spacetimes, as defined in section \ref{sec:AF}, which gives all the necessary ingredients to compute the charges, namely, the background metric $g_{ab}$, given by equation \eqref{AF} and the symmetry generators $\xi^a$, given by equation \eqref{BMSgen}.  In this case, 
\begin{equation} \label{measure}
 (d^2x)_{ab}\, \sqrt{-g} = d\Omega\ r^2 e^{2\beta} \delta_{[a}^{u} \delta^{r}_{b]}.
\end{equation}

Plugging in the above expressions into equation \eqref{AsympCharge} leads to a rather complicated expression of the form
\begin{equation} \label{BMScharge:gen}
 \ndelta \mathcal{Q}_\xi[\delta g, g]= \frac{1}{16 \pi G} \int_{S}\, d\Omega\ \Big\{ \ndelta \mathcal{I}_0 + \frac{\ndelta \mathcal{I}_1}{r} + \frac{\ndelta \mathcal{I}_2}{r^2} + \frac{\ndelta \mathcal{I}_3}{r^3} + o(r^{-3}) \Big\}.
\end{equation}
The first term $\ndelta \mathcal{I}_0$ in the expansion above has been derived in Ref.\ \cite{BarTro}, as we shall review below.  Strictly, only this first term is defined at null infinity.  Therefore, a definition of asymptotical flatness along the lines of Geroch \cite{geroch} would simply not identify any further terms beyond the leading one, $\ndelta \mathcal{I}_0$.  However, there is no reason why one should not consider the subleading terms and as we shall find below, this provides a direct relation between subleading ``BMS charges'' and the non-linear NP charges.

\subsection{BMS charge at $O(r^{0})$} \label{sec:I0}

Barnich and Troessaert \cite{BarTro} found that
\begin{equation} \label{I0}
 \ndelta \mathcal{I}_0 = \delta \big( -2 s F_{0} \big) + \frac{s}{2} \partial_u C_{IJ} \delta C^{IJ}.
\end{equation}
Significantly, the BMS charge is not integrable.  This non-integrability is directly related to the flux of gravitational radiation, or ``Bondi news,'' at null infinity \cite{BB}.  The first term on the left-hand side, $-2 s F_{0}$, would 
be a conserved charge if there were no flux at infinity.  
$-2 F_0$ is generally known as the Bondi mass aspect, and if $s$ is chosen from the $\ell=0$ or $\ell=1$ 
spherical harmonics, the charge corresponds to the Bondi-Sachs 4-momentum vector.

It should be emphasised that the above separation into an integrable and non-integrable part is not unique.  One could simply rearrange the terms differently, by moving some portion of the integrable part into the non-integrable part.  However, the most significant aspect of the above exercise is that the BMS charge at leading order is non-integrable, and that this is related to the news at null infinity.  In fact, one could ask whether the non-integrable part in equation \eqref{I0} can ever be set to zero for non-trivial parameter $s$.  Clearly, this is only possible if and only if
\begin{equation}
 \partial_u C_{IJ} = 0.
\end{equation}
This corresponds precisely to the absence of Bondi news at null infinity.

\subsection{BMS charge at $O(r^{-1})$} \label{sec:I1}

At the next order, a rather long but straightforward calculation gives that\footnote{Given equation \eqref{BMScharge:gen}, i.e.\ the fact that we always regard these quantities as being integrated over a round 2-sphere, we freely use integration by parts, ignoring total derivative terms.}$^,$\footnote{We note that there exist many Schouten identities that allow the terms to be written in different forms, see appendix \ref{app:iden}.  For example, it can be shown that (see appendix \ref{app:iden}) $$   D^{I}C^{JK} D_{I} C_{JK} - D^{I}C^{JK} D_{K} C_{IJ} - D^I C_{IK} D_J C^{JK} = 0.$$}
\begin{equation} \label{I1}
  \ndelta \mathcal{I}_1 = s \delta \left(-2 F_1 - D_I C_1^I + \frac{3}{16} (\Box - 2) C^2 +  D^{I} C_{IK} D_J C^{JK} -\frac{1}{4} D^{I}C^{JK} D_{I} C_{JK} \right).
\end{equation}
Thus, at this order the BMS charge is integrable. Moreover, from equation \eqref{F1}, we find that if the energy-momentum tensor component $T_{01} = o(r^{-4}),$ the Einstein equation implies that
\begin{equation}
 \mathcal{I}_1 = 0.
\end{equation}
If, on the other hand, $T_{01}$ is non-vanishing at this order, we have a new non-linear BMS charge
\begin{equation}
 \mathcal{Q}_1 = \int_{S}\, d\Omega\ \big(-s\, T_{01}|_{r^{-4}}\big).
\end{equation}

\subsection{BMS charge at $O(r^{-2})$} \label{sec:I2}

Similarly, at the next order, we find that
\begin{align}
 \ndelta \mathcal{I}_2 = s\ \delta  \Big( & -2 F_2 -2 D_I C_2^{I} -3 D^I C_{IJ} C_1^J -\frac{3}{2} C_{IJ} D^I C_1^J + \frac{1}{8} C^2 \, D_I D_J C^{IJ} \notag\\
  & - \frac{1}{32} C^{IJ}\, D_I D_J C^2 - \frac{1}{8}  C^{IJ} D_I C^{KL} D_J C_{KL} + \frac{3}{16} D_I C^{IJ} D_J C^{2} \Big) \notag \\[2mm]
  & \hspace{-10mm} + s \Bigg( \frac{1}{2} \Big[ \partial_u D_{IJ} \delta C^{IJ} + \delta D_{IJ} \partial_u C^{IJ} \Big] - \frac{1}{16} \partial_u C^2 \delta C^2 + \frac{1}{8} F_0 \delta C^2 - \frac{1}{2} D^I C_1^J \delta C_{IJ}  \notag \\
  &  - C_{1}^I D^J \delta C_{IJ} + \frac{1}{16} D_I D_J C^{IJ} \delta C^2 + \frac{1}{32} D_I D_J C^2 \delta C^{IJ} +\frac{1}{16} D_I C^2 D_J \delta C^{IJ} \notag \\
  & \hspace{30mm} + \frac{1}{2}C_{KL} D_I C^{IK} D_{J} \delta C^{JL}+ \frac{1}{8} \delta C^{IJ} D_I C^{KL} D_J C_{KL}  \Bigg). \label{BMScharge:I2}
\end{align}
Assuming that
\begin{equation}
 T_{0m} = o(r^{-5}), \qquad T_{01} = o(r^{-4}), \qquad T_{mm} = o(r^{-4}),
\end{equation}
which give equations for $C_2^I$ (equation \eqref{C2}), $F_2$ (equation \eqref{F2}) and $\partial_u D_{IJ}$ (equation \eqref{uD}), respectively, the expression for $\ndelta \mathcal{I}_2$ reduces to\footnote{For brevity, we have not directly substituted equation \eqref{uD} for $\del_u D_{IJ}$ into the expression below.}
\begin{align}
 \ndelta \mathcal{I}_2 =&  s\  D_I D_J \delta \Big( -  D^{IJ} + \frac{1}{16} \,  C^2 C^{IJ}\Big) \notag \\[2mm]
  &  + s \Bigg( \frac{1}{2} \Big[ \partial_u D_{IJ} \delta C^{IJ} + \delta D_{IJ} \partial_u C^{IJ} \Big] - \frac{1}{16} \partial_u C^2 \delta C^2 + \frac{1}{8} F_0 \delta C^2 - \frac{1}{2} D^I C_1^J \delta C_{IJ}  \notag \\
  & \hspace{10mm} - C_{1}^I D^J \delta C_{IJ} + \frac{1}{16} D_I D_J C^{IJ} \delta C^2 + \frac{1}{32} D_I D_J C^2 \delta C^{IJ} +\frac{1}{16} D_I C^2 D_J \delta C^{IJ} \notag \\
  & \hspace{40mm}  + \frac{1}{2}C_{KL} D_I C^{IK} D_{J} \delta C^{JL}+ \frac{1}{8} \delta C^{IJ} D_I C^{KL} D_J C_{KL}  \Bigg). \label{BMScharge:I2Einstein}
\end{align}
Thus, at order $r^{-2}$, we have a situation that is analogous to the leading BMS charge.  That is, for a general parameter $s$ there is a non-zero integrable piece as well as a non-zero non-integrable piece, presumably again related to a flux.  However, given that the expressions above do not exist \textit{at} null infinity as the boundary of the conformally compactified spacetime, the relation to quantities at null infinity is lost.  Physically the best way to think about these quantities is perhaps that they are defined ``close'' to null infinity.  For this reason we say that the non-integrable part is related to \textit{fake news} at null infinity.  While, the physical interpretation of the leading order BMS charge is clear, this is not the case here.  Of course, there is also the issue of the non-uniqueness of the split between the integral and non-integrable terms as explained before.  It will become clear later why we have chosen the above splitting.

We have established that at $O(r^{-2})$, we have a subleading BMS charge that is non-integrable for a general parameter $s$.  It is reasonable to consider whether there exists an integrable BMS charge at this order for some special parameter(s).  Given that there are no Einstein equations for $F_{0}$, $C_1^I$ and $C_{IJ}$, terms in $\ndelta \mathcal{I}_2^{(non-int)}$ involving these quantities would then have to vanish independently.   Consider first the terms involving $F_0$ in the non-integrable part in equation \eqref{BMScharge:I2}.  Using the equations for the supertranslation variations of the metric components listed in section \eqref{sec:BMS} and the Einstein equations \eqref{var:D} and \eqref{uD}, we find that the only terms in $\ndelta \mathcal{I}_2^{(non-int)}$ that contribute to terms involving $F_0$ are 
\begin{equation}
 \ndelta \mathcal{I}_2^{(non-int)}|_{F_0 \; \textrm{terms}} = s \Bigg( \frac{1}{2} \Big[ \partial_u D_{IJ} \delta C^{IJ} + \delta D_{IJ} \partial_u C^{IJ} \Big] + \frac{1}{8} F_0 \delta C^2 \Bigg)\Bigg|_{F_0 \; \textrm{terms}} .
\end{equation}
Thus, using equations \eqref{var:C}, \eqref{var:D} and \eqref{uD}
\begin{equation}
 \ndelta \mathcal{I}_2^{(non-int)}|_{F_0 \;\textrm{terms}}  = - \frac{1}{4} s F_0 C^{IJ} D_I D_J s.
\end{equation}
In order for the above term to be zero for an arbitrary symmetric, trace-free matrix $C_{IJ}$, we conclude that
\begin{equation} \label{I2:s}
  D_I D_J s = \frac{1}{2} \omega_{IJ} \Box s,
\end{equation}
i.e.\ $s$ is an $\ell=0$ or $\ell=1$ spherical harmonic, with
\begin{equation}
 \Box s = - \ell (\ell+1) s, \qquad \ell \in \{0,1\}.
\end{equation}

Next, consider the terms involving $C_1^I$.  Analogously, we find here that the only relevant terms that can contribute are
\begin{equation}
 \ndelta \mathcal{I}_2^{(non-int)}|_{C_1^I \; \textrm{terms}}  = s \Bigg( \frac{1}{2} \Big[ \partial_u D_{IJ} \delta C^{IJ} + \delta D_{IJ} \partial_u C^{IJ} \Big] -\frac{1}{2} D^I C_1^J \delta C_{IJ} - C_{1}^I D^J \delta C_{IJ} \Bigg)\Bigg|_{C_1^I \; \textrm{terms}}.
\end{equation}
Note that substituting equation \eqref{I2:s} in the variation of $C_{IJ}$ \eqref{var:C} gives that
\begin{equation} \label{s01dC}
 \delta C_{IJ} = s \partial_u C_{IJ}.
\end{equation}
Furthermore, using equations \eqref{var:D} and \eqref{uD}, we find that the terms involving $C_1^I$ then simplify to
\begin{equation}
 \ndelta \mathcal{I}_2^{(non-int)}|_{C_1^I \; \textrm{terms}} = - D_I (s\,  C_{1\, J} \delta C^{IJ}),
\end{equation}
which is a total derivative term and can thus be ignored.  

Lastly, the only terms left to consider are those involving only $C_{IJ}$.  Using equation \eqref{s01dC}, the only contributing terms are
\begin{align}
  \ndelta \mathcal{I}_2&^{(non-int)} = \frac{1}{16} s D_I C^2 D_J \delta C^{IJ} + \frac{1}{2} s C_{KL} D_I C^{IK} D_{J} \delta C^{JL} + \big(\delta D_{IJ} - s \partial_u D_{IJ} \big)\delta C^{IJ}
  \notag \\[1mm]
  &+s \Big[ \partial_u D_{IJ} - \frac{1}{8} C_{IJ} \partial_u C^2 + \frac{1}{8} C_{IJ} D_K D_L C^{KL} + \frac{1}{32} D_I D_J C^2+ \frac{1}{8} D_I C^{KL} D_J C_{KL}  \Big] \delta C^{IJ}.
\end{align} 
Substituting the $C_{IJ}$ terms in $\delta D_{IJ}$ and $\partial_u D_{IJ}$ from equations \eqref{var:D} and \eqref{uD}, respectively, and using equation \eqref{I2:s}, gives
\begin{equation}
  \ndelta \mathcal{I}_2^{(non-int)} = D_I \Big( \Big[ \frac{1}{16} s D_J C^2 + \frac{1}{2} s C_{JK} D_L C^{KL} \Big] \delta C^{IJ} \Big),
\end{equation}
i.e.\ it reduces to a total derivative, which vanishes when integrated over the
2-sphere.  Hence, we conclude that for $s$ an $\ell=0$ or $\ell=1$ spherical harmonic,
\begin{equation}
  \ndelta \mathcal{I}_2^{(non-int)} = 0.
\end{equation}
Therefore, $\ndelta \mathcal{I}_2$ is now integrable and hence we can read off the (unintegrated) charge from equation \eqref{BMScharge:I2Einstein}
\begin{equation} \label{I2:int}
 \mathcal{I}_2 =  s\  D_I D_J  \Big( -  D^{IJ} + \frac{1}{16} \,  C^2 C^{IJ}\Big).
\end{equation}
Up to total derivatives, the charge at this order is equivalently obtained by integrating
\begin{equation}
 \mathcal{I}_2 =  D_I D_J s\   \Big( -  D^{IJ} + \frac{1}{16} \,  C^2 C^{IJ}\Big).
\end{equation}
Equation \eqref{I2:s} and the trace-free property of $C_{IJ}$ and $D_{IJ}$ then implies that in fact
\begin{equation}
 \mathcal{I}_2 = 0.
\end{equation}
In conclusion, there is no non-trivial integrable charge at this order.  This result is similar in spirit to that obtained at the previous order, where we found that, while integrable, $\mathcal{I}_1 = 0$ if we assume strong enough fall-off conditions for the matter fields.

\subsection{BMS charge at $O(r^{-3})$} \label{sec:I3}

Finally, we consider the next subleading term, which we shall later relate to the NP charges in section \ref{sec:BMSNP}.  A long but straightforward calculation gives that
\begin{align}
 \ndelta \mathcal{I}_3 = s\ \delta  \Big( & -2 F_3 -3 D_I C_3^I + 2\Box \beta_2 + 4\beta_2 +\frac{3}{2} C_1^I C_{1\, I} +\frac{3}{8} D_I (C^2  C_1^I) - \frac{3}{256} (C^2)^2 \notag \\
 & - \frac{1}{2} C^{IJ} \, \Box D_{IJ} + \frac{1}{2} D^{IJ} \Box C_{IJ} + \frac{1}{2} D^I D^{JK} (4 D_K C_{IJ} - 3  D_I C_{JK}) +\frac{3}{2} D^{IJ} C_{IJ} \notag\\
  & + \frac{3}{512} \Box (C^2)^2 +\frac{13}{512} D^I C^2 D_I C^2 + \frac{1}{64} C^2 D^IC^{JK} (3 D_I C_{JK} - 4 D_K C_{IJ})  \Big) \notag\\[2mm]
  & \hspace{-10mm} + s \Bigg( \frac{1}{2} \Big[ \partial_u E_{IJ} \delta C^{IJ} + \delta E_{IJ} \partial_u C^{IJ} \Big] + \frac{1}{8} F_1 \delta C^2 - 2 C^{IJ} D_I \delta C_{2\, J} - 4 \delta C_2^I D^J C_{IJ}  \notag \\
  &  - 3 \delta C^{IJ} D_I C_{2\, J}- 5  C_2^I D^J \delta C_{IJ} - \frac{3}{4} C^2 D_I \delta C_1^I - \frac{17}{16} \delta C^2 D_I C_1^I - \frac{3}{2} \delta C_1^I D_I C^2 \notag \\
  &  - \frac{15}{8} C_1^I D_I \delta C^2 +  \frac{1}{2} C^{IJ} \delta C_{JK} D_I C_1^K + \frac{5}{2} C_1^K C^{IJ} D_I \delta C_{JK} + C_1^K  \delta C^{IJ} D_I C_{JK}   \notag \\
  &  + \frac{1}{2} C_1^K \delta C^{IJ} D_K C_{IJ} + \frac{3}{2} \delta C_1^K  C^{IJ} D_I C_{JK}  + \frac{5}{4} \delta C^{IJ} \Box D_{IJ}  + \frac{3}{4} C^{IJ} \Box \delta D_{IJ}   \notag \\  
  &  + \frac{5}{8} D^{IJ} \Box \delta C_{IJ} - \frac{11}{4} D^I D^{JK} D_K \delta C_{IJ} + \frac{15}{4} D^I D^{JK} D_I \delta C_{JK} + 3 D^J C_{IJ} D_K \delta D^{IK} \notag \\
  &  - \frac{3}{4} D^{IJ} \delta C_{IJ} - \frac{3}{2} \delta D^{IJ} C_{IJ} - \frac{1}{16} C^2 \Box \delta C^2  - \frac{3}{256} \delta C^2 \Box C^2  - \frac{1}{4} \delta C^{IJ} C_{JK} D^K D_I C^2 \notag \\
  &  - \frac{1}{32} C^2 \delta C^{IJ} D^K D_I C_{JK}  - \frac{3}{64} C^2 C^{IJ} D^K D_I \delta C_{JK} + \frac{1}{32} \delta C^2 C^{IJ} D_I D^K C_{JK} \notag \\
  &  + \frac{9}{64} C^2 D_I C^{IK} D^J \delta C_{JK} - \frac{7}{32} C^{IJ} D^K C_{JK} D_I \delta C^2 - \frac{1}{8} C^{IJ} D_I C_{JK} D^K \delta C^2 \notag \\
  &  + \frac{1}{64} \delta C^2 D^I C^{JK} D_I \delta C_{JK}  - \frac{9}{64} \delta C^{IJ} D_{I} C^2 D^K C_{JK} - \frac{17}{64} \delta C^{IJ} D^K C^2 D_{I} C_{JK} \notag \\
  &  - \frac{9}{32} C^{IJ} D_I C^2  D^{K} \delta C_{JK}  - \frac{17}{64} C^{IJ} D^K C^2 D_{I} \delta C_{JK} - \frac{7}{128} C^2 \delta C^2  \Bigg). \label{BMScharge:I3}
\end{align}
Assuming that
\begin{equation}
 T_{00} = o(r^{-6}), \qquad T_{0m} = o(r^{-6}), \qquad T_{01} = o(r^{-6}), \qquad T_{mm} = o(r^{-5}),
\end{equation}
we obtain equations for $\beta_2$ \eqref{b2}, $C_2^I$ \eqref{C2}, $C_3^I$ \eqref{C3}, $F_3$ \eqref{F3} and $\partial_u E_{IJ}$ \eqref{uE}, respectively.  Inserting these equations into \eqref{BMScharge:I3} gives the much simpler expression
\begin{align}
 \ndelta \mathcal{I}_3 = s\ \delta  \Big(& - D_I D_J E^{IJ} +\frac{1}{2} \Box (D^{IJ} C_{IJ}) - \frac{1}{32} \Box (C^2)^2 \Big) \notag\\[2mm]
  & \hspace{-10mm} + s \Bigg( \frac{1}{2} \Big[ \partial_u E_{IJ} \delta C^{IJ} + \delta E_{IJ} \partial_u C^{IJ} \Big] - \frac{1}{4} D_I (C_1^K C^{IJ}) \delta C_{JK} + \frac{1}{4} C_1^K  C^{IJ} D_I \delta C_{JK} \notag \\
  &  + \frac{1}{4} \delta C^{IJ} D^K D_I D_{JK}  + \frac{5}{4} D_{JK} D_I D^K \delta C^{IJ}  + D_I D_{JK} D^K \delta C^{IJ} \notag \\ 
  &  + \frac{1}{16} \delta C^{IJ} D^K (C_{JK} D_I C^2) -\frac{5}{64}\Big[ \delta C^{IJ} D^K (C^2 D_I C_{JK}) + C_{JK} D_I (C^2 D^K \delta C^{IJ}) \Big] \notag \\
   &  - \frac{1}{16} C_{JK} D_I C^2 D^K \delta C^{IJ}  \Bigg). \label{BMScharge:I3Einstein}
\end{align}
In deriving this equation from \eqref{BMScharge:I3}, simple applications of the identity \eqref{iden:hadi} are required, as well as the fact that the covariant derivatives in the round 2-sphere metric satisfy
\begin{equation} \label{Riemann:S2}
 [D_I, D_J] V_K = R_{IJK}{}^L V_{L}, \qquad R_{IJKL} = \omega_{IK}\, \omega_{JL} - \omega_{IL}\, \omega_{JK}.
\end{equation}

As with $\ndelta \mathcal{I}_2$ in section \ref{sec:I2}, we find that in general there exist non-integrable terms.  As before, one may consider whether there exists some choice or choices of the parameter $s$ such that the non-integrable part of $\ndelta \mathcal{I}_3$ vanishes.  We note that there are no Einstein equations for $F_0$, $C_1^I$, $D_{IJ}$ or $C_{IJ}$,  and therefore we can consider terms involving each one of these fields in isolation, without loss of generality.  

First, consider terms involving $F_0$.  Inspecting equation \eqref{BMScharge:I3Einstein} and equations \eqref{var:E} and \eqref{uE}, we find that the only terms containing $F_0$ are
\begin{align}
\ndelta \mathcal{I}_3^{(non-int)}|_{F_0 \; \textrm{terms}} &= \frac{1}{2} s \Big[ \partial_u E_{IJ} \delta C^{IJ} + \delta E_{IJ} \partial_u C^{IJ} \Big]\Big|_{F_0 \; \textrm{terms}}  \notag \\
                                  &= - \frac{1}{16} s C^2 F_0\; \omega_{IJ} \Big[ \delta C^{IJ} + s \partial_u C^{IJ} \Big].
\end{align}
Since $C_{IJ}$ is trace-free, it follows that the terms involving $F_0$ 
vanish.

Next, we consider terms involving $C_1^I$.  These come from
\begin{align}
 \ndelta \mathcal{I}_3^{(non-int)}|_{C_1^I \; \textrm{terms}} &= s \Bigg( \frac{1}{2} \Big[ \partial_u E_{IJ} \delta C^{IJ} + \delta E_{IJ} \partial_u C^{IJ} \Big] - \frac{1}{4} D_I (C_1^K C^{IJ}) \delta C_{JK} \notag \\
 & \hspace{70mm}+ \frac{1}{4} C_1^K  C^{IJ} D_I \delta C_{JK} \Bigg)\Bigg|_{C_1^I \; \textrm{terms}}  \notag \\[2mm]
 &= \frac{1}{4} D_K \Big( s C_1^I C^{JK} \delta C_{IJ} \Big) \notag \\
 & \hspace{10mm} + \frac{1}{2} \Big( D^I D^J s - \frac{1}{2} \Box s \, \omega^{IJ} \Big) \Big[ D_{K} \big(s C_1^K C_{IJ} \big) -  D_{I} \big(s C_1^K C_{JK} \big) \Big],
\end{align}
where we have used equations \eqref{var:E}, \eqref{uE} and \eqref{var:C}.  Notice that the first term in the final equation above is a total derivative and can therefore be ignored. Furthermore, up to total derivatives, the second set of 
terms is equivalent to
\begin{equation}
  \ndelta \mathcal{I}_3^{(non-int)} = - \frac{1}{2} s C_1^K C_{IJ} \Big( D_K D^I D^J s - \frac{1}{2} \delta^J_K D^I \Box s - \delta^J_K D^I s \Big),
\end{equation}
where we have made use of equation \eqref{Riemann:S2}.  Now, if this expression is to vanish for arbitrary $C_1^K$ and symmetric trace-free $C_{IJ}$, the symmetrisation on $(IJ)$ of the terms in the bracket would need to be proportional to the round 2-sphere metric $\omega_{IJ}$.  Contracting over the $IJ$ indices determines the function of proportionality.  In summary, we find that $s$ must satisfy
\begin{equation} \label{I3:s}
 D_K D_{(I} D_{J)} s - \frac{1}{2} \omega_{K(I} D_{J)} \Box s - \frac{1}{4} \omega_{IJ} D_{K} \Box s - \omega_{K(I} D_{J)} s + \frac{1}{2} \omega_{IJ} D_{K} s = 0.
\end{equation}
As discussed in appendix C, this equation is satisfied if $s$ is any $\ell=2$ spherical harmonic (see equation \eqref{l2:1}).  In particular,
\begin{equation} \label{boxs}
 \Box s = - 6 s,
\end{equation}
and equation \eqref{I3:s} reduces to the simpler equation \eqref{l2id2}
\be
D_K D_I D_J s = -2 \, \omega_{IJ}\, D_K\,s -
    2 \, \omega_{K(I}\, D_{J)}\, s .\,\label{l2id2text}
\ee

Assuming henceforth that $s$ is an $\ell = 2$ spherical harmonic, we proceed to investigate the terms featuring $D_{IJ}$, which appear in the following terms
\begin{align}
  \ndelta \mathcal{I}_3^{(non-int)}|_{D_{IJ} \; \textrm{terms}} &= s \Bigg( \frac{1}{2} \Big[ \partial_u E_{IJ} \delta C^{IJ} + \delta E_{IJ} \partial_u C^{IJ} \Big] + \frac{1}{4} D^K D_I D_{JK} \delta C^{IJ} \notag \\
  & \hspace{28mm}  + \frac{5}{4} D_{JK} D_I D^K \delta C^{IJ}+ D_I D_{JK} D^K \delta C^{IJ} \Bigg)\Bigg|_{D_{IJ} \; \textrm{terms}}  \notag \\[2mm]
 &= \frac{5}{4} D_I \Big( s D_{JK} D^K \delta C_{IJ} \Big) + D^K \Big(s \delta C^{IJ} D_I D_{JK} - \frac{5}{4} \delta C^{IJ} D_I \big( s D_{JK} \big) \Big) \notag \\
 & \hspace{25mm} - \frac{1}{2} \Big( D^I D^J s - \frac{1}{2} \Box s \, \omega^{IJ} \Big) D^K \Big[ s D_I D_{JK} +5 D_{JK} D_I s \Big], 
\end{align}
where, as before, we have used equations \eqref{var:E}, \eqref{uE} and \eqref{var:C}.  The first two terms in the final equation here are total derivatives, and 
so when integrated over the sphere they will give zero. Up to total derivatives, the remaining terms then give
\begin{equation} \label{I3:Dterms}
  \ndelta \mathcal{I}_3^{(non-int)} = \frac{1}{2} D^K \Big( D^I D^J s - \frac{1}{2} \Box s \, \omega^{IJ} \Big) \Big[ s D_I D_{JK} +5 D_{JK} D_I s \Big].
\end{equation}
Using equations \eqref{boxs} and \eqref{l2id2text}, one can show that
\begin{equation} \label{s:iden}
 D^K \Big( D^I D^J s - \frac{1}{2} \Box s \, \omega^{IJ} \Big) = 2 \omega^{I[J} D^{K]} s - \omega^{JK} D^I s.
\end{equation}
Given that the above combination is contracted with terms that are symmetric and trace-free in $(JK)$ in equation \eqref{I3:Dterms}, this implies that the terms involving $D_{IJ}$ vanish in $\ndelta \mathcal{I}_3^{(non-int)}$.

Finally, we are left with terms involving only $C_{IJ}$
\begin{align}
  \ndelta \mathcal{I}_3^{(non-int)} &= s \Bigg( \frac{1}{2} \Big[ \partial_u E_{IJ} \delta C^{IJ} + \delta E_{IJ} \partial_u C^{IJ} \Big] + \frac{1}{16} \delta C^{IJ} D^K (C_{JK} D_I C^2)  \notag \\
  & \hspace{2mm} -\frac{5}{64}\Big[ \delta C^{IJ} D^K (C^2 D_I C_{JK}) + C_{JK} D_I (C^2 D^K \delta C^{IJ}) \Big]  - \frac{1}{16} C_{JK} D_I C^2 D^K \delta C^{IJ} \Bigg)  \notag \\[2mm]
 &= \frac{5}{64} D^K \Big(C^2 D_I(s C_{JK}) \delta C^{IJ} \Big) -\frac{5}{64} D_I \Big(s C^2 C_{JK} D^K \delta C^{IJ} \Big)  \notag \\
 & \hspace{83mm} -\frac{1}{16} D^K \Big(s C_{JK} D_I C^2 \delta C^{IJ} \Big) \notag \\
 & \quad  - \frac{1}{8} \Big( D^I D^J s - \frac{1}{2} \Box s \, \omega^{IJ} \Big) D^K \Big[ s C_{JK} D_I C^2 - \frac{5}{4} C^2 D_{I}(s C_{JK}) \Big], 
\end{align}
where we have used equations \eqref{var:E}, \eqref{uE} and \eqref{var:C}.  Up to total derivatives,
\begin{equation} \label{I3:Cterms}
  \ndelta \mathcal{I}_3^{(non-int)} = \frac{1}{8} D^K \Big( D^I D^J s - \frac{1}{2} \Box s \, \omega^{IJ} \Big) \Big[ s C_{JK} D_I C^2 - \frac{5}{4} C^2 D_{I}(s C_{JK}) \Big].
\end{equation}
Equation \eqref{s:iden}, and the fact that $C_{JK}$ is symmetric and trace-free, then imply that
\begin{equation}
  \ndelta \mathcal{I}_3^{(non-int)} = 0.
\end{equation}

In summary, we find that the non-integrable terms in $\ndelta \mathcal{I}_3$ vanish if and only if $s$ is an $\ell=2$ spherical harmonic.  Thus, we have an integrable charge, whose integrand can be read off from equation \eqref{BMScharge:I3Einstein}.  Using equation \eqref{trE}, this gives, for any $\ell=2$ spherical
harmonic $s$,
\begin{equation} \label{I3:int}
 \mathcal{I}_3 = s\, D_I D_J \Bigg( -E^{IJ} + \frac{1}{2}\, \textup{tr}E\ \omega^{IJ} \Bigg),
\end{equation}
which, up to total derivatives, is equivalent to 
\begin{equation}
 \mathcal{I}_3 = - \left( D_I D_J s + 3 s\, \omega_{IJ} \right) E^{IJ},
\end{equation}
where we have used equation \eqref{boxs}.
Hence, we have found a new integrable charge that is generally non-vanishing for arbitrary field $E_{IJ}$.  In the next section, we shall demonstrate that this charge has a precise correspondence with the NP charges.

\section{Relating the BMS charges to the NP formalism} \label{sec:BMSNP}

In this section, we relate the tower of BMS charges found in section \ref{sec:BMSsub} to the formalism developed by Newman and Penrose in Ref.\ \cite{NP61, NP}.  In particular, we show that the BMS charges at order $r^{-3}$ are the non-linear NP charges discovered in Ref.\ \cite{NP}. Throughout this section, we use the notation of the Newman-Penrose formalism, which can be found in Ref.\ \cite{NP61}.\footnote{In Ref.\ \cite{NP61}, they use negative signature convention, whereas we use positive signature conventions.  This simply means that the scalar product of the null frame vectors and the definition of the Newman-Penrose scalars is different by a minus sign. }

The Newman-Penrose formalism begins with a choice of complex null frame $\{\ell,n,m,\bar{m}\}$.  We choose the null frame defined in equation \eqref{AF:frame}.  Once a null frame has been chosen, we can form scalars by contracting tensors onto null frame components.  Hencewith, 12 complex spin coefficients are formed by contracting covariant derivatives of the null frame vectors onto null frame components.  The spin coefficients constitute information about the connection.  For example,
\begin{equation}
 \kappa = m^a \ell^b \nabla_b \ell_a, \qquad \sigma = -m^a m^b \nabla_b \ell_a
\end{equation}
parameterise geodesicity and shear, respectively, of the null vector congruence associated with $\ell$.  Moreover, we have scalars representing the ten degrees of freedom in the Ricci tensor, and the five complex Weyl scalars
\begin{gather}
 \Psi_0 = \ell^a m^b \ell^c m^d C_{abcd}, \quad \Psi_1 = \ell^a n^b \ell^c m^d C_{abcd}, \quad \Psi_2 = \ell^a m^b \bar{m}^c n^d C_{abcd}, \notag \\
 \Psi_3 = \ell^a n^b \bar{m}^c n^d C_{abcd}, \quad  \Psi_4 = n^a \bar{m}^b n^c \bar{m}^d C_{abcd}.
\end{gather}

With the fall-off conditions \eqref{met:falloff} and \eqref{falloff:matter}, we find that
\begin{gather}
 \Psi_0 = \psi_0^0\; \frac{1}{r^5} + \psi_0^1\; \frac{1}{r^6} + o(r^{-6}), \quad \Psi_1 = \psi_1^0\; \frac{1}{r^4} + o(r^{-4}), \quad \Psi_2 = \psi_2^0\; \frac{1}{r^3} + \psi_2^1\; \frac{1}{r^4} + o(r^{-4}), \notag \\[2mm]
 \Psi_3 = \psi_3^0\; \frac{1}{r^2} + o(r^{-2}), \quad \Psi_4 = \psi_4^0\; \frac{1}{r} + o(r^{-1}).
\end{gather}
The above property of the Weyl tensors is known as peeling \cite{NP61, bondi, sachs}.  Moreover,
\begin{equation}
 \sigma = \sigma^0\; \frac{1}{r^2} + o(r^{-2}).
\end{equation}
In terms of the functions that define the metric components \eqref{met:falloff} and \eqref{def:fg},
\begin{equation}
 \sigma^0 = \frac{(1+i)}{2} (f_0+ig_0).
\end{equation}
Defining the differential operators $\eth$ and $\bar{\eth}$ acting on a scalar of spin $n$ \cite{Goldberg:1966uu, NP61}\footnote{The spins $n$ of the Weyl scalars $\Psi_0$, $\Psi_1$, $\Psi_2$, $\Psi_3$, $\Psi_4$ are 2, 1, 0, -1 and -2, respectively, while $\sigma$ has spin 2.  Complex conjugation reverses the sign of the spin: $n \rightarrow -n$. }
\begin{align}
 \eth \eta &= - \frac{(1+i)}{2}\sin^n \theta  \left( \frac{\partial }{\partial \theta }-\frac{i}{\sin\theta} \frac{\partial}{\partial \phi }\right)\Big(\frac{\eta}{\sin ^n\theta}\Big), \notag \\[2mm]
 \bar{\eth} \eta &= - \frac{(1-i)}{2} \frac{1}{\sin^n \theta}  \left( \frac{\partial }{\partial \theta }+\frac{i}{\sin\theta} \frac{\partial}{\partial \phi }\right)\big(\sin ^n\theta\, \eta\big),
\end{align}
\begin{equation} \label{NPeqns}
 \psi_4^0 = - \partial_u^2 \bar{\sigma}^0, \quad \psi_3^0 =  \eth \partial_u\bar{\sigma}^0, \quad \psi_2^0 - \bar{\psi}_2^0 = \bar{\sigma}^0 \partial_u \sigma^0 -  \sigma^0 \partial_u \bar{\sigma}^0 + \bar{\eth}^2 \sigma^0 - \eth^2 \bar{\sigma}^0.
\end{equation}
Furthermore,
\begin{equation} \label{psi20}
 \psi_2^0 + \bar{\psi}_2^0 = F_0 - \partial_u |\sigma^0|^2
\end{equation}
and\footnote{Note that $(C_1^{\theta} - i \sin \theta\, C_1^{\phi})$ is a spin 1 quantity.}
\begin{align}
 \psi_2^1 &= F_1 + \frac{(1+i)}{2}\, \bar{\eth}(C_1^{\theta} - i \sin \theta\, C_1^{\phi}) - \frac{(1-i)}{4}\, \eth(C_1^{\theta} + i \sin \theta\, C_1^{\phi}) \notag \\
            & \quad - \frac{3}{4} \eth(\bar{\sigma}^0 \bar{\eth} \sigma^0) + \frac{9}{4} \sigma^0 \bar{\eth} \eth \bar{\sigma}^0 + \frac{1}{4} \bar{\eth} \bar{\sigma}^0 \eth \sigma^0, \\
 \psi_1^0 &= \frac{3(1+i)}{4} (C_1^{\theta} - i \sin \theta\, C_1^{\phi}) + \frac{3}{4}  \eth |\sigma^0|^2 + 3 \sigma^0 \eth \bar{\sigma}^0, \\
 \psi_0^0 &= -3 (1+i) (f_2 + i g_2) - i (f_0^3 + g_0^3 ) + \frac{(1-i)}{4} (f_0+ig_0)^3, \\
 \psi_0^1 &= -6 (1+i) (f_3 + i g_3).
\end{align}

Now that we have defined all the quantities in the language of Newman and Penrose we are ready to compare to the tower of BMS charges derived in section \ref{sec:BMSsub}.

\subsection{$\mathcal{I}_0$ and BMS charges}
The standard BMS charge is defined by
\begin{equation} \label{BMS:charge}
 P_{\ell,m} = - \frac{1}{2\pi G} \int d\Omega\ Y_{\ell m}\; (\psi_2^0 + \sigma^0 \partial_u \bar{\sigma}^0),
\end{equation}
where $Y_{\ell m}$ are the usual spherical harmonics.
Setting $0 \leq |m| \leq \ell \leq 1$ gives the usual Bondi-Sachs 4-momentum vector.  In fact, in this case, from the last equation in \eqref{NPeqns}
\begin{equation}
 \Im(\psi_2^0 + \sigma^0 \partial_u \bar{\sigma}^0) = \Im(\bar{\eth}^2 \sigma^0)
\end{equation}
is a total derivative.  Thus,
\begin{equation} \label{BMS:charge01}
 P_{\ell,m} = - \frac{1}{2\pi G} \int d\Omega\ Y_{\ell m}\ \Re(\psi_2^0 + \sigma^0 \partial_u \bar{\sigma}^0), \qquad \ell\in \{0,1\}.
\end{equation}

%
%

Defining the integrable part of equation \eqref{I0} to be
\begin{equation}
 \mathcal{Q}_0 = \frac{1}{8\pi G} \int d\Omega\ Y_{\ell m} (-2 F_0)
\end{equation}
with $s=Y_{\ell m}$ and rewriting the above expression in terms of Newman-Penrose quantities gives
\begin{equation} \label{Q0}
 \mathcal{Q}_0 = - \frac{1}{2\pi G} \int d\Omega\ Y_{\ell m}\ \Re(\psi_2^0 + \sigma^0 \partial_u \bar{\sigma}^0) 
\end{equation}
Comparing with equation \eqref{BMS:charge} we find that the charge above is the real part of the BMS charge as defined by Newman-Penrose (see equation (4.15) of Ref.\ \cite{NP}).  However, for $\ell=0,1$, they are equal as can be seen from equation \eqref{BMS:charge01}.

The integrability property of $\mathcal{Q}_0$ in the language of Barnich-Brandt translates to its conservation along null infinity in the language of Newman-Penrose.  The Bianchi identities, which are non-trivial in the Newman-Penrose formalism, imply that
\begin{equation}
 \partial_u \psi_2^0 = -  \eth^2\partial_u \bar{\sigma}^0 - \sigma^0 \partial_u^2 \bar{\sigma}^0.
\end{equation}
Using this equation
\begin{equation}
\partial_u(-2 F_0) = - 4 \partial_u \Re(\psi_2^0 + \sigma^0 \partial_u \bar{\sigma}^0) = \Re(\eth^2\partial_u \bar{\sigma}^0) -4 |\partial_u \sigma^0|^2.
\end{equation}
Note that for $\ell \leq 1$, the first term is a total derivative since\footnote{This result comes from standard properties of spin-weighted spherical harmonics (see e.g.\ Ref.\ \cite{NP}) \label{ft:spin}
\begin{equation*}
 \eth(_{s}Y_{lm}) = \sqrt{(l-s)(l+s+1)}\ _{s+1}Y_{lm}, \qquad
  \bar{\eth}(_{s}Y_{lm}) = - \sqrt{(l+s)(l-s+1)}\ _{s-1}Y_{lm}.
\end{equation*}
}
\begin{equation} \label{eth20}
 \bar{\eth}^2 Y_{\ell m} = \eth^2 Y_{\ell m} = 0,
\end{equation}
i.e.\ it is a soft graviton term \cite{Strominger:2013jfa
}, while in terms of functions of the metric components
\begin{equation}
 |\partial_u \sigma^0|^2 = \frac{1}{8} \partial_u C_{IJ} \partial_u C^{IJ},
\end{equation}
i.e.\ the obstacle to the conservation of $\mathcal{Q}_0$ is
\begin{equation}
 \frac{1}{2} \partial_u C_{IJ} \partial_u C^{IJ},
\end{equation}
which matches precisely with the non-integrable term in equation \eqref{I0}.

\subsection{$\mathcal{I}_1$ and $\psi_1^0$}
Writing $\mathcal{I}_1$ from equation \eqref{I1} in terms of Newman-Penrose quantities gives
\begin{equation}
 \mathcal{I}_1 = 2\, \Re (\bar{\eth} \psi_1^0 - \psi_2^1).
\end{equation}
The Bianchi identities imply that 
\begin{equation}
 \psi_2^1 = \bar{\eth} \psi_1^0.
\end{equation}
Hence, 
\begin{equation}
 \mathcal{I}_1 = 0.
\end{equation}

\subsection{$\mathcal{I}_2$ and $\psi_0^0$}
In section \ref{sec:I2}, we found that choosing $s$ to be an $\ell=0$ or $\ell=1$ mode, the non-integrable part vanishes and we are left with a candidate charge of the form \eqref{I2:int}.  In terms of Newman-Penrose quantities,
\begin{equation} 
  D_I D_J  \Big( -  D^{IJ} + \frac{1}{16} \,  C^2 C^{IJ}\Big) = \frac{2}{3} \Re \big( \bar{\eth}^2 \psi_0^0 \big).
\end{equation}
Hence,
\begin{equation} 
 \mathcal{I}_2 =  \frac{2}{3}\ Y_{\ell m}\ \Re \big( \bar{\eth}^2 \psi_0^0 \big)
\end{equation}
with $\ell=0,1$.  Using equation \eqref{eth20}, we reproduce the result in section \ref{sec:I2} that the integrable charge is in fact zero.

\subsection{$\mathcal{I}_3$ and NP charges}
In section \ref{sec:I3}, we found an integrable charge at order $r^{-3}$ as long as $s$ is chosen to be an $\ell=2$ spherical harmonic.  Translating the main result of that section, equation \eqref{I3:int}, into Newman-Penrose language, and using the fact that
\begin{equation} \label{Epsi01}
 D_I D_J \Bigg( -E^{IJ} + \frac{1}{2} \omega^{IJ} \Big[ D^{KL} C_{KL} - \frac{1}{16} (C^2)^2 \Big] \Bigg) = \frac{1}{3} \Re \big( \bar{\eth}^2 \psi_0^1 \big),
\end{equation}
gives
\begin{equation} 
 \mathcal{Q}_3 = \frac{1}{24\pi G} \int d\Omega\ \bar{Y}_{2,m} \Re \big( \bar{\eth}^2 \psi_0^1 \big).
\end{equation}
Integrating by parts gives
\begin{equation}
 \mathcal{Q}_3 = \frac{1}{4\sqrt{6}\, \pi G } \int d\Omega\ \Big[ {}_{2}\bar{Y}_{2,m}\ \psi_0^1 + (-1)^m\ {}_{2}Y_{2,-m}\ \bar{\psi_0^1} \Big].
\end{equation}
Notice that the first term in the integrand above corresponds to the NP charges (see equation (4.19) of Ref.\ \cite{NP}).  The second term is not quite the complex conjugate of the first.  However, the combination means that we only have half the number of NP charges.  Perhaps an easier way to see this is that in equation \eqref{Epsi01}, only the real part of $\bar{\eth}^2 \psi_0^1$ appears on the right-hand side.

\section{Discussion} \label{sec:dis}

In this paper, we have established concretely the relation of the NP charges to the BMS group of asymptotic symmetries at null infinity and its associated 
charges.  While the relation of the NP charges to the BMS group was argued for in Ref.\ \cite{NP}, even an explicit demonstration of the supertranslation invariance of the non-linear NP charges has been missing (see, however, Ref.\ \cite{goldberg}).  In particular, interestingly, we find that the NP charges appear 
at subleading $1/r^3$ order in a $1/r$-expansion of the Barnich-Brandt charge, which defines the standard BMS charge at leading order.  

We have used the Barnich-Brandt definition of asymptotic charges, but this is not unique.  For example, the Iyer-Wald definition \cite{IW} differs by a term of the form 
\begin{equation} 
 \frac{1}{16 \pi G} \int_{S}\,(d^2x)_{ab}\, \sqrt{-g}\ \big( \nabla^a \xi^c + \nabla^c \xi^a \big) g^{bd} \delta g_{cd}.
\end{equation}
In fact, as discussed in Ref.~\cite{compere}, the above expression, with an arbitrary coefficient, represents a one parameter family of ambiguities.  Our results in this paper are not affected by the inclusion of this term.

Curiously, we only obtain half the number of NP charges, owing to the fact that the Barnich-Brandt charge is real.  It would be interesting to understand whether the Barnich-Brandt integral could ever give all ten NP charges and, if so, how.  
It seems unlikely that the SL$(2,\mathbb{C})$ part, or indeed its 
generalisation involving superrotations, could account for the remaining
five charges.

Another slightly puzzling feature of the Barnich-Brandt charge definition is 
that in it $s$ plays the role both of the supertranslation parameter and also as a function used in order to define the charge.  Thus, for example, in section \ref{sec:I3}, when we show that $\mathcal{I}_3$ is integrable if $s$ is 
an $\ell=2$ harmonic, showing that the variation of $\mathcal{I}_3$ with such a parameter 
$s$ vanishes clearly does not prove that the integrable charge is invariant under the full action of the full supertranslation group.  Rather, it only demonstrates that $\mathcal{I}_3$ is invariant under the action of those supertranslations where the supertranslation parameter $s$ is an $\ell=2$ harmonic.  We do, 
however, prove the complete invariance of the NP charges under the full action of the supertranslation group in appendix \ref{app:STNP}.

At the linearised level, at each order in the $1/r$ expansion, there are conserved charges associated to the tower of linearised Newman-Penrose charges. Conde and Mao~\cite{conde} also find only half of these charges, \emph{viz.}\ the real parts. Linearising our extended BMS charges, at each order we get the same form as Conde-Mao's charges. At suitably low enough order the Conde-Mao charges come from expanding $F(u,r,\theta,\phi)$, which we also have. Therefore, at leading order our charges agree; see equation \eqref{I0}. However, at subleading orders, we also get contributions from the expansion of $D_{I} C^{I}(u,r,\theta,\phi)$; see equations \eqref{I1}, \eqref{BMScharge:I2} and \eqref{BMScharge:I3}. Using equations of motion \eqref{F1}, \eqref{F2}, \eqref{C2}, \eqref{F3} and \eqref{C3}, the Taylor coefficients in the $1/r$ expansion of $F(u,r,\theta,\phi)$ and $D_{I} C^{I}(u,r,\theta,\phi)$ are proportional to each other, hence the form of our linearised charges at each order is equal to the charges of Conde-Mao. However, the coefficients are different. In particular, at the subleading order the relative constant of proportionality between $F_1(u,\theta,\phi)$ and $D_{I} C_1^{I}(u,\theta,\phi)$ is such that they cancel upon use of equation \eqref{F1}. The difference between our and Conde-Mao's linearised charges reflects the fact that at the linearised level there are a number of independent supertranslation invariant quantities. However, at the non-linear level this degeneracy is lifted and there is a unique combination that is supertranslation invariant, which is what is found in this paper.

The fact that there are only ten non-linearly conserved NP charges has not been fully understood in the context of the Newman-Penrose formalism.  It remains an open question whether the reframing of the charges in terms of the Barnich-Brandt formalism could help with resolving this puzzle.  Of course, a prerequisite to understanding this is first to understand why half the NP charges are missing in this formalism.

In a future work, we will also investigate the tower of subleading BMS charges for the more realistic fall-off conditions at infinity~\cite{Angelopoulos:2016wcv, Angelopoulos:2017iop,Angelopoulos:2018uwb} that do not preclude some physical processes, such as compact data close to spacelike infinity.  These fall-off conditions are most relevant for current gravitational wave observations and the hope would be that this leads to the discovery of a quantity that is useful for gravitational wave analysis.

It would also be interesting to investigate the charge algebra at subleading order. In particular, there will be a hierarchy of BMS algebras at each order with different modified brackets, corresponding to the different fake news at each order, and field-dependent central extensions. At the leading order, the algebra has no central extension for supertranslation generators \cite{BarTro} and this is expected to be the case at subleading orders as well. However, extending our charges to include rotations should give rise to new central extensions at subleading orders. Furthermore, at $O(1/r^3),$ there ought to be a subalgebra, given by the generators corresponding to the Newman-Penrose charges, for which the modified bracket is just given by the ordinary Dirac bracket. We will investigate the charge algebra hierarchy in a future work. 

\section*{Acknowledgements}

We would like to thank Gary Gibbons, Pujian Mao, Blagoje Oblak, 
Malcolm Perry, Shahin Sheikh-Jabbari and C\'edric Troessaert for useful discussions.  We would like to thank the Mitchell Family Foundation for hospitality at the Brinsop Court workshop.  
Moreover, M.G.\ and C.N.P.\ would like to thank the Max-Planck-Institut f\"ur Gravitationsphysik (Albert-Einstein-Insitut), Potsdam, where this work was initiated, H.G.\ would like to thank the Mitchell Institute for Fundamental Physics and Astronomy, Texas A\&M University and H.G.\ and M.G.\ would like to thank the ICTP, Trieste for hospitality during the course of this work.
M.G.\ is partially supported by grant no.\ 615203 from the European Research Council under the FP7.  C.N.P.\ is partially supported by DOE grant DE-FG02-13ER42020.

\appendix

\section{Supertranslation invariance of NP charges} \label{app:STNP}

In this appendix, we demonstrate the supertranslation invariance of the NP charges in the language of Ref.\ \cite{NP}.  The ten non-linear NP charges are given in terms of $\psi_0^1$,
\begin{equation} \label{NPcharges}
 G_m = \int d\Omega\  {}_2\bar{Y}_{2,m}\, \psi_0^1,
\end{equation}
 where
\begin{equation}
 \psi_0 = \psi_0^0\, r^{-5} + \psi_0^1\, r^{-6} + o(r^{-6}).
\end{equation}

We would like to investigate the effect of a supertranslation on the NP charges.  Note that in terms of the metric components
\begin{align} 
 \psi_0^0 &= -3 (1+i) (f_2 + i g_2) - i \left(f_0^3+g_0^3\right) + \frac{(1-i)}{4}  (f_0 + i g_0)^3, \label{psi00} \\
 \psi_0^1 &= -6 (1+i) (f_3 + i g_3).  \label{psi01}
\end{align}
Using the expression for $\psi_0^1$ above and equations \eqref{var:E}, \eqref{var:D} and \eqref{var:C}, a straightforward yet slightly cumbersome calculation shows that
\begin{equation} \label{ST:psi01}
 \delta \psi_0^1 = s\, \partial_u \psi_0^1 - 5\,  \bar{\eth} \left( \psi_0^0\; \eth s \right) - \bar{\eth} s\; \eth \psi_0^0  + 4 \bar{\eth} s\, \sigma^0 \psi_1^0.
\end{equation}
Now, equation (4.12) of Ref.\ \cite{NP} reads\footnote{There is in fact a minor typographical error in equation (4.12) of Ref.\ \cite{NP}.} 
\begin{equation}
\partial_u \psi_0^1 = - s \bar{\eth} \left(\eth \psi_0^0 - 4 \sigma^0 \psi_1^0\right).
\end{equation}
Substituting the above equation in equation \eqref{ST:psi01} gives
\begin{equation}
 \delta \psi_0^1 = - \bar{\eth}\left( s\; \eth \psi_0^0  + 4 \bar{\eth} s\, \sigma^0 \psi_1^0\right) - 5\,  \bar{\eth} \left( \psi_0^0\; \eth s \right) .
\end{equation}
Therefore, from equation \eqref{NPcharges}, the change of the NP charges under the action of a supertranslation generator is
\begin{equation}
  \delta G_m = - \int d \Omega\ {}_2\bar{Y}_{2,m}\, \bar{\eth}\left( s\; \eth \psi_0^0  + 4 \bar{\eth} s\, \sigma^0 \psi_1^0 + 5\,  \psi_0^0\; \eth s \right)
\end{equation}
Using the fact that 
\begin{equation}
 \bar{\eth}\, {}_2\bar{Y}_{2,m}=0
\end{equation}
the expression above reduces to a total derivative.  Thus, 
\begin{equation}
 \delta  G_m = 0,
\end{equation}
i.e.\ we conclude that the NP charges are invariant under supertranslations.

\section{Identities for tensors on the 2-sphere} \label{app:iden}

For the calculation in section \ref{sec:BMSsub}, it is useful to be aware of a number of identities satisfied by tensors on the 2-sphere.  These are ultimately derived from Schouten identities in two dimensions.  For example, for any symmetric traceless matrix $X_{IJ}$, such as $C_{IJ}$ or $D_{IJ}$,
\begin{equation} \label{iden:chris}
 X_{IJ} = - \epsilon_{IK} \epsilon_{JL} X^{KL},
\end{equation}
where $\epsilon_{IJ}$ is the volume 2-form on the round 2-sphere.  This can be derived from the fact that
\begin{equation} \label{iden:ee}
 \epsilon_{IK} \epsilon_{JL} = 2 \, \omega_{I[J} \omega_{L]K}.
\end{equation}
Now, consider 
\begin{align}
 X_{IJ} \delta^L_K &=  - \epsilon_{IM} \epsilon_{JN} X^{MN} \epsilon^{LP} \epsilon_{KP} \notag \\
                   &=  - \epsilon_{JN} \epsilon_{KP} (\delta^L_I X^{PN} - \delta^P_I X^{LN}) \notag \\
                   &=  \delta^L_I X_{JK} + \omega_{JK} X^L{}_{I} - \omega_{IJ} X^L{}_K,
\end{align}
where we have used equation \eqref{iden:chris} in the first equality and \eqref{iden:ee} a number of times in the calculation above.  Hence, we derive the 2-dimensional Fierz identity
\begin{equation} \label{iden:fierz}
  \omega_{IJ} X_{KL} + \omega_{KL} X_{IJ} - \omega_{IL} X_{JK} - \omega_{JK} X_{IL}=0,
\end{equation}
or contracting this equation with an arbitrary $V^{L}$
\begin{equation} \label{iden:hadi}
 X_{IJ} V_K = X_{JK} V_I + X_{IL} V^L \omega_{JK} - X_{KL} V^L \omega_{IJ}.
\end{equation}
Applying the above identity to the first three indices in $X_{KI} X_J{}^K$ gives
\begin{align}
 X_{KI} X_J{}^K &= X_{IJ} X_K{}^K + X^2 \omega_{IJ} - X_{KI} X_J{}^K,
\end{align}
where $X^2 = X_{IJ} X^{IJ}.$
Using the fact that $X_{IJ}$ is trace-free implies that
\begin{equation}
 X_{IK} X^{JK} = \frac{1}{2} X^2 \delta_I^J. 
\end{equation}
Similarly,
\begin{align}
  D^{I} X^{JK} D_{I}X_{JK} &= D^{I} X^{JK} (D_K X_{IJ} + D^L X_{KL} \omega_{IJ} - D^L X_{LI} \omega_{JK}) \notag \\
                           &= D^I X^{JK} D_K X_{IJ} + D^I X_{IK} D_J X^{JK}      
\end{align}
or
\begin{equation}
   D^{I}X^{JK} D_{I} X_{JK} - D^{I}X^{JK} D_{K} X_{IJ} - D^I X_{IK} D_J X^{JK} = 0.
\end{equation}
One may derive many other equations from identity \eqref{iden:fierz} or equivalently \eqref{iden:hadi} in a similar fashion to the derivations above.

\section{$\ell=0$, $\ell=1$ and $\ell=2$ spherical harmonics} \label{app:l2}

The spherical harmonics $Y_{\ell m}(\theta,\phi)$ on the unit 2-sphere obey
$-\square Y_{\ell m}= \ell(\ell+1) Y_{\ell m}$.  The $\ell=0$ harmonic is of course just a constant.

Suppose $\psi$ is an $\ell=1$ harmonic, satisfying $-\square\psi=2\psi$.  It
follows that $\psi$ satisfies the equation
\be
D_I D_J \psi =\ft12\omega_{IJ}\, \square \psi\,,\label{l1id}
\ee
where $\omega_{IJ}$ is the unit 2-sphere metric, 
and hence
\be
  D_I D_J \psi = -\omega_{IJ}\, \psi\,.\label{l1id2}
\ee
One can prove (\ref{l1id}) by defining $T_{IJ}\equiv 
D_I D_J \psi -\ft12\omega_{IJ}\, \square \psi$, and observing that, after
integrating $|T_{IJ}|^2\equiv T^{IJ}\, T_{IJ}$ 
over the sphere and performing some integrations by parts,
\be
\int |T_{IJ}|^2 d\Omega = \ft12 \int \psi \square (\square+2)\psi\, d\Omega
\,.
\ee
Thus if $\psi$ obeys $-\square\psi=2\psi$ then $T_{IJ}$ must vanish,
hence establishing (\ref{l1id}).\footnote{Of course if $\square\psi=0$ then
$T_{IJ}$ again vanishes and (\ref{l1id}) also holds, but trivially in this
case since $\psi$ is then a constant.}

   Turning now to $\ell=2$ modes, let us define the tensor
\be
T_{IJK} \equiv D_K D_I D_J\psi - \ft13 \omega_{IJ}\, D_K\,\square\psi
  -\ft13 \omega_{K(I}\, D_{J)}\, \square\psi\,.
\ee
Integrating $|T_{IJK}|^2 \equiv T^{IJK} T_{IJK}$ over the sphere and
performing some integrations by parts, we find
\be
\int |T_{IJK}|^2 \, d\Omega = -\fft5{18} \int\psi \square(\square+6)
    (\square +\ft{12}{5})\psi\, d\Omega\,.\label{TIJK}
\ee
Thus, if $\psi$ is an $\ell=2$ harmonic, meaning that it satisfies
$-\square\psi=6\psi$, then $T_{IJK}=0$ and so\footnote{An $\ell=0$
mode (a constant) also trivially satisfies (\ref{l2id}).   A function 
satisfying $-\square\psi = \ft{12}{5}\psi$ cannot be smooth 
on $S^2$ and would violate the assumptions under which (\ref{TIJK})
was derived.  Thus, such a function does not obey (\ref{l2id}).}
\be
D_K D_I D_J\psi =\fft13 \omega_{IJ}\, D_K\,\square\psi +
    \fft13 \omega_{K(I}\, D_{J)}\, \square\psi\,.\label{l2id}
\ee
It follows also that it obeys
\be
D_K D_I D_J\psi = -2\,\omega_{IJ}\, D_K\,\psi -
    2 \,\omega_{K(I}\, D_{J)}\,\psi\,\label{l2id2}
\ee
and
\begin{equation} \label{l2:1}
 D_K D_{(I} D_{J)} s = \frac{1}{2} \omega_{K(I} D_{J)} \Box s + \frac{1}{4} \omega_{IJ} D_{K} \Box s + \omega_{K(I} D_{J)} s - \frac{1}{2} \omega_{IJ} D_{K} s.
\end{equation}

  It is interesting to note that the above identities generalise to
higher dimensions, and here we record these for the case of a unit $n$-sphere.
The analogous hyperspherical harmonics have eigenvalues $-\square =
\ell(\ell+n-1)$.  The $\ell=1$ modes $\psi$, obeying $-\square\psi=n\,\psi$,
satisfy
\be
D_I D_J \psi = \fft1{n}\, \omega_{IJ}\, \square\psi = -\omega_{IJ}\, \psi\,,
\label{l1ndim}
\ee
and the $\ell=2$ modes, obeying $-\square\psi = 2(n+1)\, \psi$, satisfy
\be
D_K D_I D_J\psi =\fft1{n+1} \, \omega_{IJ}\, D_K\,\square\psi +
    \fft1{n+1}\, \omega_{K(I}\, D_{J)}\, \square\psi =
-2\omega_{IJ}\, D_K\,\psi -
    2 \omega_{K(I}\, D_{J)}\,\psi\,.\label{l2ndim}
\ee
These identities can again be proven by integrating the squares of the
analogously-defined tensors $T_{IJ}$ and $T_{IJK}$ over the sphere.

\section{Barnich-Brandt charge and the Einstein equation}
In this appendix, we show that the Barnich-Brandt charge as applied to asymptotically-flat spacetimes is zero upon use of the Einstein equations.  Starting from equation \eqref{AsympCharge} and rearranging the terms gives that
\begin{gather} 
  \ndelta \mathcal{Q}_\xi[\delta g, g]= \frac{1}{16 \pi G} \int_{S}\,(d^2x)_{ab}\, \sqrt{-g}\ \Big\{3 \big[ \xi^b g^{cd} \nabla^a \delta g_{cd} - \xi^b g^{ac} \nabla^d \delta g_{cd} 
  - g^{ad} \delta g_{cd} \nabla^b \xi^c \big] \hspace{15mm} \notag   \\[2mm]
  \hspace{60mm} +  \nabla^b \Big( g^{cd} \delta g_{cd} \xi^a + 2 g^{ad} \delta g_{cd}  \xi^c \Big) - \nabla^c (\xi^a g^{bd} \delta g_{cd}) \Big\}. \label{BB:1}
\end{gather}
Using 
\begin{equation}
 \delta g_{ab} = 2 \nabla_{(a} \xi_{b)},
\end{equation}
the above expression reduces to
\begin{gather} 
  \ndelta \mathcal{Q}_\xi[\delta g, g]= \frac{1}{16 \pi G} \int_{S}\,(d^2x)_{ab}\, \sqrt{-g}\ \Big\{3 \big[2 \xi^b \nabla^a \nabla^c \xi_{c} - \xi^b \nabla^c \nabla^a \xi_{c} - \xi^b \nabla^c \nabla_c \xi^{a}
  - \nabla^c \xi^a \nabla^b \xi_{c} \big] \hspace{5mm} \notag   \\[2mm]
  \hspace{35mm} +  \nabla^b \Big(2 \xi^a \nabla_c \xi^c + 2 \xi^c \nabla_c \xi^a + \nabla^a \xi^2 \Big) - \nabla^c (\xi^a \nabla^b \xi_c + \xi^a \nabla_c \xi^b) \Big\}, \label{BB:2}
\end{gather}
where we have also used the fact that by symmetry,
\begin{equation}
\nabla^{[a} \xi_c \nabla^{b]} \xi^{c} = 0.
\end{equation}
Now arranging the last two terms in the first line of equation \eqref{BB:2} 
above, and also using
\begin{equation}
\nabla^c \xi^{[a} \nabla_{c} \xi^{b]} = 0, \quad \nabla^{[a} \nabla^{b]} \xi^2 = 0, 
\end{equation}
gives
\begin{gather} 
  \ndelta \mathcal{Q}_\xi[\delta g, g]= \frac{1}{16 \pi G} \int_{S}\,(d^2x)_{ab}\, \sqrt{-g}\ \Big\{6 \xi^b [\nabla^a, \nabla^c] \xi_{c} +  2  \nabla^b \nabla_c (\xi^a \xi^c) \hspace{40mm} \notag   \\[2mm]
  \hspace{80mm}  - 2 \nabla_c (\xi^b \nabla^c \xi^a - 2 \xi^b \nabla^a \xi^c) \Big\}, \label{BB:3}
\end{gather}
which simplifies to
\begin{gather} 
  \ndelta \mathcal{Q}_\xi[\delta g, g]= \frac{1}{16 \pi G} \int_{S}\,(d^2x)_{ab}\, \sqrt{-g}\ \Big\{6 \xi^b [\nabla^a, \nabla^c] \xi_{c} +  2  [\nabla^b, \nabla_c] (\xi^a \xi^c) \hspace{35mm} \notag   \\[2mm]
  \hspace{80mm}  + 2 \nabla_c (\xi^c \nabla^b \xi^a + \xi^a \nabla^c \xi^b - \xi^a \nabla^b \xi^c) \Big\}. \label{BB:4}
\end{gather}
For now we ignore the terms in the first line in the equation above and focus on the terms in the second line.  In fact, we shall demonstrate that these terms form a total derivative.  Performing explicitly the contraction in $ab$ and using equation \eqref{measure}, the terms in the second line become
\begin{gather} 
 \frac{r^2}{8 \pi G} \int_{S}\,d\Omega\ e^{2\beta}\  \nabla_c \Big\{ \xi^{[c} \nabla^{r]} \xi^u + \xi^{[u} \nabla^{c]} \xi^r - \xi^{[u} \nabla^{r]} \xi^c \Big\}. \label{BB:5}
\end{gather}
The expression in the braces clearly vanishes when $c=u$ or $c=r$.  Hence, the above equation reduces to
\begin{gather} 
 \frac{r^2}{4 \pi G} \int_{S}\,d\Omega\ e^{2\beta}\  \nabla_I X^{Iur}, \label{BB:6}
\end{gather}
where
\begin{equation}
 X^{cab} = \xi^c \nabla^{[b} \xi^{a]} + \xi^{[a} \nabla^{|c|} \xi^{b]} - \xi^{[a} \nabla^{b]} \xi^c.
\end{equation}
Note that as argued above $X^{uur}=X^{rur}=0$. Now,
\begin{align}
 \nabla_I X^{Iur} &= \partial_I X^{Iur} + \Gamma^I_{IJ} X^{Jur} + \Gamma^u_{Ic} X^{Icr} + \Gamma^r_{Ic} X^{Iuc} \notag \\
                  &= \partial_I X^{Iur} + \Gamma^I_{IJ} X^{Jur} + (\Gamma^u_{Iu} + \Gamma^r_{Ir}) X^{Iur} + \Gamma^u_{IJ} X^{IJr} - \Gamma^r_{IJ} X^{IJu}.
\end{align}
From the definition of $X^{cab}$, it can be shown that
\begin{equation}
 X^{IJa} = X^{aIJ},
\end{equation}
i.e.\ that $X^{IJa} = X^{[IJ]a}$.  Hence,
\begin{equation}
 \nabla_I X^{Iur} = \partial_I X^{Iur} + (\Gamma^I_{IJ}  + \Gamma^u_{Ju} + \Gamma^r_{Jr}) X^{Jur}.
\end{equation}
Inserting the expressions for the Christoffel symbols \cite{BarnichAspects}
\begin{equation}
 \Gamma^I_{IJ} + \Gamma^u_{Iu} + \Gamma^r_{Ir} = {}^{(\omega)}\Gamma^I_{IJ} + 2 \partial_I \beta,
\end{equation}
where ${}^{(\omega)}\Gamma^I_{IJ}$ is the Christoffel symbols associated with the round 2-sphere metric $\omega_{IJ}$, equation \eqref{BB:6} simplifies to
\begin{equation}
 \frac{r^2}{4 \pi G} \int_{S}\,d\Omega\ D_I \Bigg( e^{2\beta}\,  X^{Iur} \Bigg) = 0.
\end{equation}
Thus, returning to equation \eqref{BB:4} and using the definition of the Riemann tensor
\begin{equation}
 [\nabla_a, \nabla_b] V_c = R_{abc}{}^d V_{d},
\end{equation}
we obtain
\begin{equation}
 \ndelta \mathcal{Q}_\xi[\delta g, g]= \frac{r^2}{2 \pi G} \int_{S}\,d\Omega\ e^{2\beta} \ \xi^{[u} R^{r]}{}_c \xi^c = \frac{r^2}{2 \pi G} \int_{S}\,d\Omega\ e^{2\beta} \ \xi^{[u} G^{r]}{}_c \xi^c.
\end{equation}
Hence, we find that on-shell
\begin{equation}
  \ndelta \mathcal{Q}_\xi[\delta g, g] = 0.
\end{equation}

\bibliographystyle{utphys}
\bibliography{NP}

\end{document}